\documentclass[letterpaper, 10 pt, conference]{ieeeconf}  

\IEEEoverridecommandlockouts                              

\usepackage{graphics} 
\usepackage{epsfig} 
\usepackage{amssymb}  
\usepackage{algorithm}
\usepackage{mathtools}
\usepackage{caption}
\usepackage{cancel}
\usepackage{graphicx,subfigure}
\usepackage[usenames,dvipsnames]{color}
\usepackage[nospace]{cite}
\usepackage[margin=1in]{geometry}
\usepackage{cuted}
\usepackage{lipsum}
\usepackage{accents}

\usepackage{amsmath,algorithm,tabularx}
\usepackage[noend]{algpseudocode}
\makeatletter
\newcommand{\multiline}[1]{%
	\begin{tabularx}{\dimexpr\linewidth-\ALG@thistlm}[t]{@{}X@{}}
		#1
	\end{tabularx}
}

\makeatletter
\def\mycopyright#1{%
	\protected@xdef \@thanks {\@thanks \protect \footnotetext [\the \c@footnote ]{#1}}%
	\protected@xdef \@bs@thanks {\@bs@thanks \protect \footnotetext [\the \c@footnote ]{#1}}
}
\makeatother

\graphicspath {{figures/}}

\title{\LARGE \bf Temporal phase unwrapping using deep learning}

\author{ Wei Yin$^{1,2,3}$, Qian Chen$^{1,2,7}$, Shijie Feng$^{1,2,3}$, Tianyang Tao$^{1,2,3}$, Lei Huang$^{4}$, \\
Maciej Trusiak$^{5}$, Anand Asundi$^{6}$, and Chao Zuo$^{1,2,3,*}$
    \thanks{{\footnotesize $^{1}${School of Electronic and Optical Engineering, Nanjing University of Science and Technology, No. 200 Xiaolingwei Street, Nanjing, Jiangsu Province 210094, China}}} 
    \thanks{{\footnotesize $^{2}${Jiangsu Key Laboratory of Spectral Imaging $\&$ Intelligent Sense, Nanjing University of Science and Technology, Nanjing, Jiangsu Province 210094, China}}}
     \thanks{{\footnotesize $^{3}${Smart Computational Imaging (SCI) Laboratory, Nanjing University of Science and Technology, Nanjing, Jiangsu Province 210094, China}}}
     \thanks{{\footnotesize $^{4}${Brookhaven National Laboratory, NSLS II 50 Rutherford Drive, Upton, New York 11973-5000, United States}} }
     \thanks{{\footnotesize $^{5}${Institute of Micromechanics and Photonics, Warsaw University of Technology, 8 Sw. A. Boboli Street, Warsaw 02-525, Poland}} }
     \thanks{{\footnotesize $^{6}${Centre for Optical and Laser Engineering (COLE), School of Mechanical and Aerospace Engineering, Nanyang Technological University, Singapore 639798, Singapore}}}
     \thanks{{\footnotesize $^{7}${e-mail: chenqian@njust.edu.cn}} }
     \thanks{{\footnotesize $^{*}${Corresponding author: zuochao@njust.edu.cn and surpasszuo@163.com}}}
}

\begin{document}
	\maketitle
	\thispagestyle{empty}
	\pagestyle{empty}
	
\begin{abstract}
The multi-frequency temporal phase unwrapping (MF-TPU) method, as a classical phase unwrapping algorithm for fringe projection profilometry (FPP), is capable of eliminating the phase ambiguities even in the presence of surface discontinuities or spatially isolated objects. For the simplest and most efficient case, two sets of 3-step phase-shifting fringe patterns are used: the high-frequency one is for 3D measurement and the unit-frequency one is for unwrapping the phase obtained from the high-frequency pattern set. The final measurement precision or sensitivity is determined by the number of fringes used within the high-frequency pattern, under the precondition that the phase can be successfully unwrapped without triggering the fringe order error. Consequently, in order to guarantee a reasonable unwrapping success rate, the fringe number (or period number) of the high-frequency fringe patterns is generally restricted to about 16, resulting in limited measurement accuracy. On the other hand, using additional intermediate sets of fringe patterns can unwrap the phase with higher frequency, but at the expense of a prolonged pattern sequence. Inspired by recent successes of deep learning techniques for computer vision and computational imaging, in this work, we report that the deep neural networks can learn to perform TPU after appropriate training, as called deep-learning based temporal phase unwrapping (DL-TPU), which can substantially improve the unwrapping reliability compared with MF-TPU even in the presence of different types of error sources, e.g., intensity noise, low fringe modulation, and projector nonlinearity. We further experimentally demonstrate for the first time, to our knowledge, that the high-frequency phase obtained from 64-period 3-step phase-shifting fringe patterns can be directly and reliably unwrapped from one unit-frequency phase using DL-TPU. These results highlight that challenging issues in optical metrology can be potentially overcome through machine learning, opening new avenues to design powerful and extremely accurate high-speed 3D imaging systems ubiquitous in nowadays science, industry, and multimedia.
\end{abstract}

\maketitle

\section{Introduction}
    \noindent
        Many imaging systems, such as fringe projection profilometry (FPP) \cite{gorthi2010fringe,geng2011structured,feng2018high}, optical interferometry \cite{vest1979holographic,gahagan2000measurement}, synthetic aperture radar (InSAR) \cite{bamler1998synthetic,curlander1991synthetic}, X-ray crystallography \cite{momose1995demonstration}, and magnetic resonance imaging \cite{haacke1999magnetic}, make use of the phase to produce the information about physical and geometrical properties of the measured objects. For instance, in FPP, the phase is proportional to the surface profile; in optical interferometry, the phase can be exploited to infer shape, deformation, vibration, and structure of the object’s surface; in InSAR, the phase is proportional to the terrain elevation height; in magnetic resonance imaging, the phase is used to measure temperature, to map the main magnetic field inhomogeneity, to identify veins in the tissues, and to segment water from fat. In these imaging modalities and applications, the initial phases of the measured data are obtained by a simple operation of arctangent from the raw complex images and are wrapped in the principal interval $(-\pi,\pi]$. Therefore, the wrapped phases distribute within the limits of $-\pi$ and $\pi$, and phase discontinuities occur at the limits every time when the unknown true phase changes by $2\pi$. The applications of the phase images require to unwrap the phases in order to remove the discontinuities and to obtain an estimate of the true phases \cite{su2004reliability,flynn1997two,zuo2016temporal,schofield2003fast,pritt1996phase,chavez2002understanding}.

        Over the past decades, many phase unwrapping algorithms have been proposed. In terms of the working domains, these algorithms can be classified into two groups: spatial phase unwrapping (SPU) \cite{su2004reliability,flynn1997two} and temporal phase unwrapping (TPU) \cite{zuo2016temporal}. Under the assumption of spatial continuity, SPU is based on the analysis of adjacent or neighboring pixels on an individual wrapped phase map, but it is challenging to cope with the disjoint regions and phase discontinuities. Conversely, TPU uses more than one frequency and unwrap the phase via temporal analysis of multiple wrapped phase values for each sample point separately. It can achieve pixel-wise phase unwrapping even in the presence of large phase discontinuities or spatially isolated regions. So far, there are typically three approaches to TPU: multi-frequency (hierarchical) approach (MF-TPU), multi-wavelength (heterodyne) approach, and number-theoretical approach. We have detailedly analyzed and discussed the unwrapping success rate and noise-resistance ability of these three approaches in a comparative review, revealing that the MF-TPU approach provides the highest unwrapping reliability and best noise-robustness among others \cite{zuo2016temporal}.

        The subsequent content of this paper will be focused on the MF-TPU approach, with an emphasis on the application of high-speed FPP \cite{su2010dynamic,zhang2018high}. In such a context, to improve the measurement efficiency, it is always desirable for the phase unwrapping algorithm to be as reliable as possible with the minimum number of projected patterns \cite{zhang2017robust}. With the recent developments in the area of digital display, digital video projectors have been increasingly used as the projection units of FPP systems, which are able to accurately control various attributes of the projected fringe patterns at high speed in software, facilitating the effective implementation of MF-TPU algorithm. For the simplest and most efficient case of MF-TPU, two sets of 3-step phase-shifting fringe patterns (totally 6 patterns) are used: the high-frequency one is for 3D measurement and the unit-frequency one is for unwrapping the phase obtained from the high-frequency pattern set. The final measurement precision or sensitivity is determined by the number of fringes used within the high-frequency patterns, under the precondition that the phase can be successfully unwrapped without triggering the fringe order error. Consequently, in order to guarantee a reasonable unwrapping success rate, the fringe number (or period number) of the high-frequency fringe patterns is generally restricted to about 16, resulting in limited measurement accuracy. On the other hand, using an additional intermediate set of fringe patterns (totally 3 sets of phase-shifting patterns) can unwrap the phase with higher frequency or higher success rate \cite{zhang2017robust}. As a result, the increased number of required patterns reduces the measurement efficiency of FPP, which is especially undesirable for high-speed real-time measurement applications.

        In this work, we demonstrate the use of a deep neural network to significantly enhance the performance of TPU compared with conventional MF-TPU. This learning-based framework uses only two (one unit-frequency, one high-frequency) wrapped phase maps obtained from 3-step phase-shifting fringe patterns as input, and directly outputs an unwrapped version of the same phase map with high reliability. Deep learning \cite{lecun2015deep} is a method based on the representation of data in machine learning for data analysis and prediction and have been applied to various fields such as computer vision, speech recognition, and natural language processing, where they have produced results that surpass the performance of traditional algorithms and are comparable or superior in some cases to human experts. Recently, machine learning-based methods have been further successfully applied to solving challenging problems in computational imaging, such as phase retrieval \cite{sinha2017lensless}, lensless on-chip microscopy \cite{rivenson2018phase}, fringe pattern analysis \cite{feng2019fringe}, and computational ghost imaging \cite{shimobaba2018computational,lyu2017deep}.

        Inspired by the great successes of deep learning techniques for these fields, here we adopt deep neural networks to beat the TPU problem, which can substantially improve the unwrapping reliability compared with MF-TPU even in the presence of different types of error sources. To validate the proposed approach, we recover the absolute phases of various tested objects by projecting fringe patterns with different frequencies, such as 1, 8, 16, 32, 48, and 64, all of which demonstrate the successful removal of phase unwrapping errors arising from the intensity noise, low fringe modulation, and intensity nonlinearity. We then demonstrate for the first time, to our knowledge, that the high-frequency phase obtained from 64-period 3-step phase-shifting fringe patterns can be directly and reliably unwrapped from one unit-frequency phase, facilitating high-accuracy high-speed 3D surface imaging with use of only 6 projected patterns without exploring any prior information and geometric constraint. These results highlight that machine learning is able to potentially overcome challenging issues in optical metrology, and provides new possibilities to design powerful high-speed FPP systems.

\section{Principle}
    \subsection{Phase-shifting profilometry (PSP)}
        \noindent
        In a typical FPP system, some pre-defined fringe patterns are projected onto the tested object surface by a digital video projector, and the camera captures the deformed fringe patterns from a different perspective. The phase of the distorted fringe patterns contains the depth information of the object, which can be further reconstructed through phase retrieval and optical triangulation  \cite{gorthi2010fringe,geng2011structured,zhang2006novel,du2007three,li2008accurate,huang2010least}. Many fringe analysis techniques have been proposed to extract the phase distribution from the distorted fringe(s), such as phase-shifting profilometry (PSP) \cite{zuo2018phase}, Fourier transform profilometry (FTP) \cite{takeda1983fourier,su2001fourier}, windowed Fourier transform profilometry \cite{kemao2007two}, and wavelet transform profilometry \cite{zhong2004spatial}. Among plenty of state-of-the-art techniques, the standard N-step phase-shifting algorithm, as a common phase-shifting algorithm for PSP, can achieve pixel-wise phase measurement and provide highest measurement resolution and accuracy since it is robust to the ambient illumination and varying surface reflectivities. In high-speed 3D shape measurement, the 3-step phase-shifting algorithm using the theoretical minimum number of fringe images for PSP is desirable for reducing the measurement time and motion-induced phase-shifting errors of dynamic measurement. The designed standard 3-step phase-shifting fringe patterns with shift offset of ${2\pi /3}$ in the projector space can be represented as
        \begin{equation}\label{1}
        \begin{aligned}
            & {I^p_n(x^p,y^p)} = 0.5 + 0.5\cos(2\pi f x^p - 2\pi n /3),
        \end{aligned}
        \end{equation}
        where ${{I^p_n(x^p,y^p)}(n = 0, 1, 2)}$ represent fringe patterns to be projected, ${f}$ is the frequency of fringe patterns. After projected onto the object surfaces, the deformed fringe patterns captured by the camera can be described as
        \begin{equation}\label{2}
        \begin{aligned}
            & {I^c_n(x,y)} = A(x,y) + B(x,y)\cos({\Phi(x,y)} - 2\pi n /3),
        \end{aligned}
        \end{equation}
        where ${A(x,y)}$, ${B(x,y)}$, and ${\Phi(x,y)}$ are the average intensity, the intensity modulation, and the phase distribution of the measured object. According to the least-squares algorithm, the wrapped phase ${\phi(x,y)}$ can be obtained as  \cite{srinivasan1984automated,de1995derivation,surrel1996design}:
        \begin{equation}\label{3}
        \begin{aligned}
            & {\phi(x,y)} = \tan^{-1} \frac{ \sqrt{3}({I^c_1(x,y)} - {I^c_2(x,y)})}{2{I^c_0(x,y)} - {I^c_1(x,y)} - {I^c_2(x,y)}}.
        \end{aligned}
        \end{equation}
        Due to the truncation effect of the arctangent function, the obtained phase ${\phi(x,y)}$ is wrapped within the range of $(-\pi,\pi]$, and its relationship with ${\Phi(x,y)}$ is:
        \begin{equation}\label{4}
        \begin{aligned}
            & {\Phi(x,y)} = {\phi(x,y)} + 2\pi k(x,y),
        \end{aligned}
        \end{equation}
        where ${k(x,y)}$ represents the fringe order of ${\Phi(x,y)}$, and its value range is from ${0}$ to ${N-1}$. ${N}$ is the period number of the fringe patterns (i.e., ${N = f}$). In FPP, the core challenge for the absolute phase recovery is to obtain ${k(x,y)}$ for each pixel in the phase map quickly and accurately.

    \subsection{Multi-frequency temporal phase unwrapping (MF-TPU)}
        \noindent
        In temporal phase unwrapping (TPU), the wrapped phase ${\phi(x,y)}$ is unwrapped with the aid of one (or more) additional wrapped phase map with different frequency. For instance, two wrapped phases ${\phi_h(x,y)}$ and ${\phi_l(x,y)}$ are both retrieved from phase shifting algorithm by using Eq. (\ref{3}), ranging from ${-\pi}$ to ${\pi}$. It is easy to find that the two absolute phases ${\Phi_h(x,y)}$ and ${\Phi_l(x,y)}$ corresponding to ${\phi_h(x,y)}$ and ${\phi_l(x,y)}$ have the following relationship:
        \begin{equation}\label{5}
        \left\{
        \begin{aligned}
            & \Phi_h(x,y) = {\phi_h(x,y)} + 2\pi k_h(x,y),\\
            & \Phi_l(x,y) = {\phi_l(x,y)} + 2\pi k_l(x,y),\\
            & \Phi_h(x,y) = ({f_h}/{f_l})\Phi_l(x,y),
        \end{aligned}
        \right.
        \end{equation}
        where ${f_h}$ and ${f_l}$ are the frequency of high-frequency fringes and low-frequency fringes. Based on the principle of MF-TPU, ${k_h(x,y)}$ can be calculated by the following formula:
        \begin{equation}\label{6}
        \begin{aligned}
            & {k_h(x,y)} = \frac{({f_h}/{f_l})\Phi_l(x,y) - \phi_h(x,y)}{2\pi}.
        \end{aligned}
        \end{equation}
        Since the fringe order ${k_h(x,y)}$ is integer, ranging from 0 to ${f_h}$ - 1, Eq. (\ref{6}) can be adapted as
        \begin{equation}\label{7}
        \begin{aligned}
            & {k_h(x,y)} = Round\left [ \frac{({f_h}/{f_l})\Phi_l(x,y) - \phi_h(x,y)}{2\pi} \right ],
        \end{aligned}
        \end{equation}
        where ${Round ()}$ is the rounding operation. When ${f_l}$ is 1, there will be no phase ambiguity so that ${\Phi_l(x,y)}$ is inherently an unwrapped phase. Theoretically, for MF-TPU, this single-period phase can be to directly assist phase unwrapping of ${{\phi_h(x,y)}}$ with relatively higher frequency. However, the phase unwrapping capability of MF-TPU is greatly constrained due to the influence of noise in practice. Assuming phase errors in the wrapped phase maps ${\phi _h(x,y)}$ and ${\Phi _l(x,y)}$ are ${\Delta \phi _h(x,y)}$ and ${\Delta \phi _l(x,y)}$ respectively, from Eq. (\ref{6}) we have:
        \begin{equation}\label{8}
        \begin{aligned}
            \Delta k(x,y) = \frac{({f_h}/{f_l})\Delta \phi_l(x,y) - \Delta \phi_h(x,y)}{2\pi},
        \end{aligned}
        \end{equation}
        Let ${\Delta \phi _{max} = max(|{\Delta \phi _h(x,y)}|,|{\Delta \phi _l(x,y)}|)}$, from Eq. (\ref{8}) we can find the upper bound of ${\Delta k(x,y)}$:
        \begin{equation}\label{9}
        \begin{aligned}
            \Delta k_{max}(x,y) = |\frac{({f_h}/{f_l})\Delta \phi_l(x,y) - \Delta \phi_h(x,y)}{2\pi}| \\
            = \Delta \phi _{max}\frac{{f_h} + {f_l}}{2\pi {f_l}}.
        \end{aligned}
        \end{equation}
        To avoid errors in determining the fringe orders, from Eqs. (\ref{7}) and (\ref{9}) we have:
        \begin{equation}\label{10}
        \begin{aligned}
            \Delta k_{max}(x,y) = \Delta \phi _{max}\frac{{f_h} + {f_l}}{2\pi {f_l}} < 0.5.
        \end{aligned}
        \end{equation}
        Subsequently, we can confirm the boundary of $\Delta \phi_{max}(x,y)$:
        \begin{equation}\label{11}
        \begin{aligned}
             0 \le  \Delta \phi_{max}(x,y) < \frac{\pi {f_l}}{{f_h} + {f_l}}.
        \end{aligned}
        \end{equation}
        Notably, Eq. (\ref{11}) defines the range of ${\Delta \phi _{max}}$ where the absolute phase can be correctly recovered. Otherwise, error will occur in determining the exact ${k_h(x,y)}$. In MF-TPU, since the frequency of the low-frequency fringes is fixed to 1, it can be found from Eq. (\ref{11}) that the higher the frequency of the high-frequency fringes, the narrower the range of ${\Delta \phi _{max}}$, and the worse the reliability of the phase unwrapping. Consequently, for a normal FPP system, MF-TPU can only reliably unwrap the phase with about 16 periods due to the nonnegligible noises and other error sources in actual measurement. Thus, it generally exploits multiple (\textgreater 2) sets of phases with different frequencies to hierarchically unwrap the wrapped phase step by step, and finally arrives at the absolute phase with high frequency instead of only using the phase with a single period. Obviously, MF-TPU, which consumes additional time for projecting patterns with intermediate frequencies, is not a good choice to realize high-speed, high-precision 3D shape measurement based on FPP.
        \begin{figure*}[htbp]
            \centerline{\includegraphics[width=2\columnwidth]{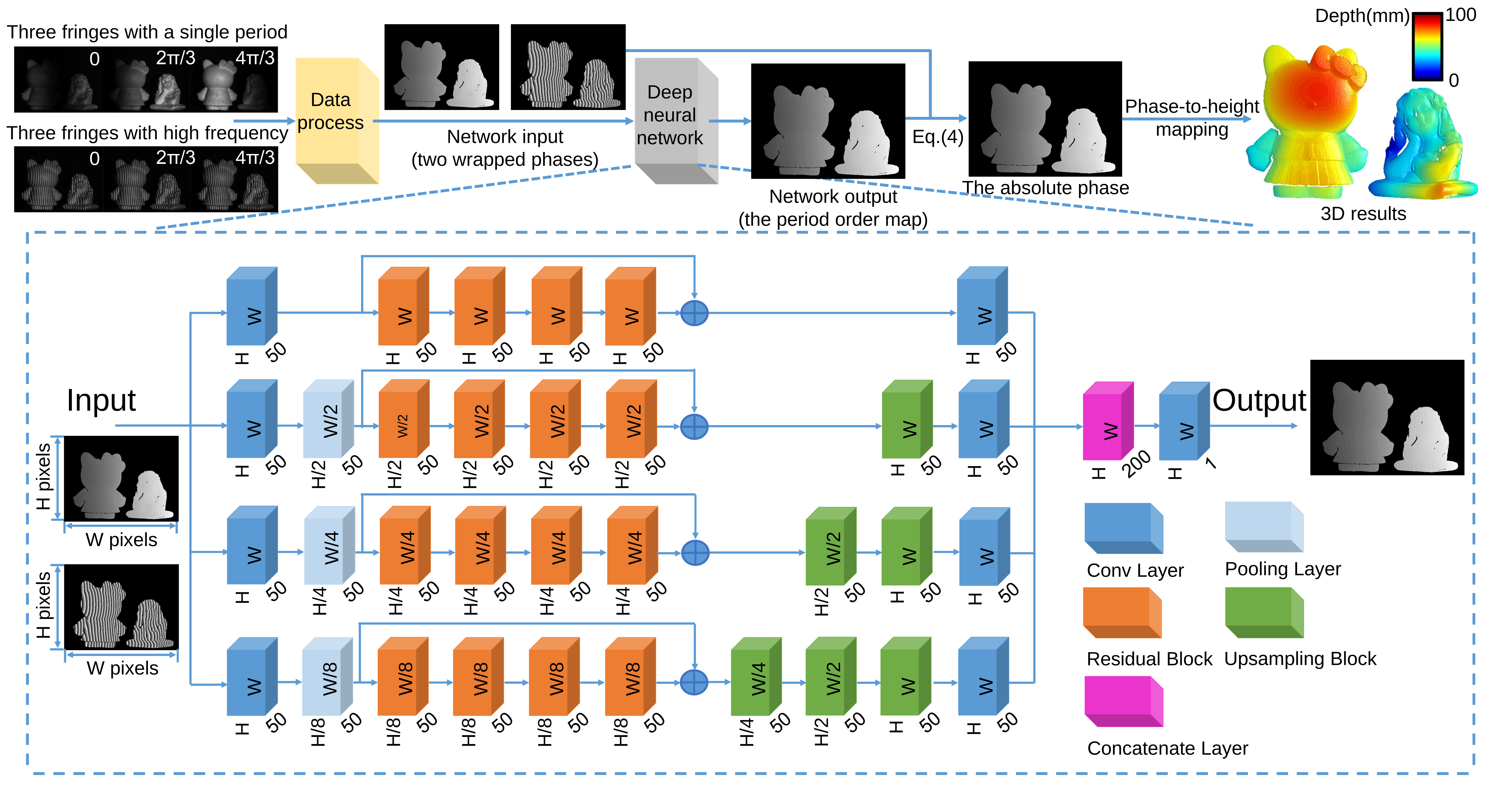}}
            \caption{
            The diagram of the proposed method. The whole framework is composed of data process, deep neural network, and phase-to-height mapping. Data process is performed to extract phases and remove the background from fringe images according to Eq. (\ref{3}) and supplementary Equation S1. Deep neural network, consisting of convolutional layers, pooling layers, residual blocks, upsampling blocks, and concatenate layer, is used to predict the period order map ${k_h(x,y)}$ from the input data (${\Phi_l(x,y)}$ and ${\phi_h(x,y)}$). Then, using Eq. (\ref{4}), ${\Phi_h(x,y)}$ is obtained and converted into 3D results after phase-to-height mapping.
            }
            \label{Fig1}
        \end{figure*}

     \subsection{Deep-learning based temporal phase unwrapping (DL-TPU)}
        \noindent
        Aiming at this problem, we choose to use the deep neural networks (DNN) to overcome the limitations of MF-TPU, and the specific diagram of the proposed method is shown as in Fig. \ref{Fig1}. The input data of the network are the two wrapped phases of the single period and high frequency, which is the same as the two-frequency TPU. To realize the highest unwrapping reliability, we adopt the residual network as the basic skeleton of our neural network \cite{he2016deep}, which can speed up the convergence of deep networks and improve network performance by adding layers with considerable depth. Then, we introduce the multi-scale pooling layer to down-sampling the input tensors, which can compress and extract the main features of the tensors for reducing the computation complexity and preventing the over-fitting. Correspondingly, it is inconsistent for the tensors sizes in the different paths after the processing of the pooling layer. Therefore, different numbers of upsampling blocks will be used to make the sizes of the tensors in the respective paths uniform (see Supplementary  Section 1 for details) \cite{shi2016real}. Moreover, it has been found that implementing shortcuts between residual blocks contributes to making the convergence of the network more stable. Although the purpose of building the network is to achieve phase unwrapping and obtain the absolute phase, there is no need to directly set the absolute phase as the network's label. Since ${\Phi_h(x,y)}$ is simply the linear combination of ${k_h(x,y)}$ and ${\phi_h(x,y)}$, according to Eq. (\ref{4}), ${\Phi_h(x,y)}$ can be obtained immediately if ${k_h(x,y)}$ is known. Once ${k_h(x,y)}$ is set as the output data of the network, it is easy to understand that the complexity of the network will be greatly reduced so that the loss of the network will converge faster and more stable, and the prediction accuracy of the network is effectively improved. It should be noted that different from the traditional SPU and TPU that the phase unwrapping is performed by utilizing the phase information solely in the spatial or temporal domain, the proposed method based on deep neural network is able to learn feature extraction and data screening, thus can exploit the phase information in the spatial and temporal domain simultaneously, providing more degrees of freedom and possibilities to achieve significantly better unwrapping performance (refer to Supplementary Section 3 for details).

        Then, using Eq. (\ref{4}), ${\Phi_h(x,y)}$ is obtained and converted into 3D results after phase-to-height mapping. In preparation for phase-to-height mapping, the projection matrices of the camera and projector need to be obtained through system calibration \cite{li2008accurate,zhang2000flexible}. Besides, in order to speed up the reconstruction, we suggest phase-to-height mapping to be implemented with a graphics processing unit \cite{feng2015graphics} or several look-up tables \cite{liu2010dual}, which can greatly save the time cost of the 3D reconstruction.

\section{Experiment results}
    \subsection{Quantitative comparison with MF-TPU}
\begin{figure}[htbp]
            \centerline{\includegraphics[width=1\columnwidth]{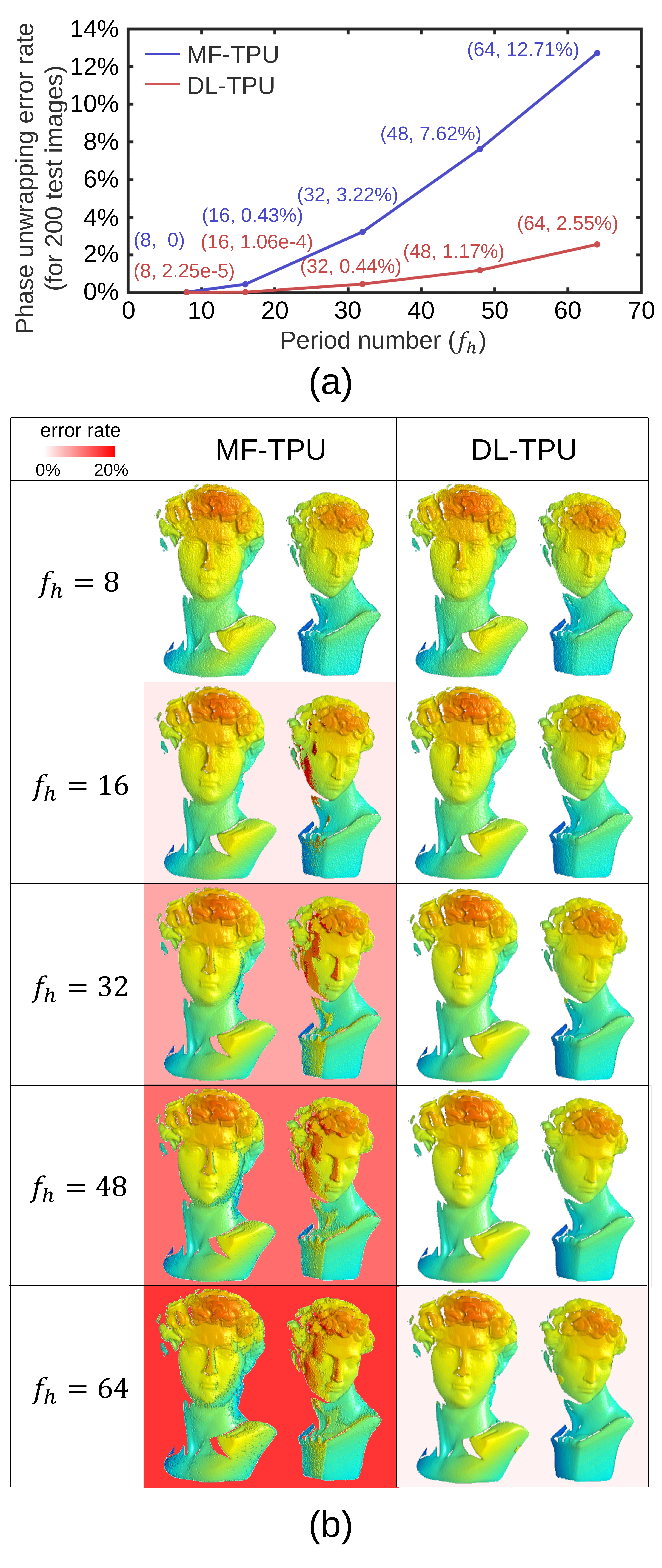}}
            \caption{
            (a) Comparison of the average error rates of phase unwrapping with different high frequencies (such as 8, 16, 32, 48 and 64) on the testing dataset using MF-TPU and DL-TPU. (b) Comparison of the 3D reconstruction results after phase unwrapping with different high frequencies (such as 8, 16, 32, 48 and 64) for a representative sample on the testing dataset using MF-TPU and DL-TPU.
            }
            \label{Fig2}
        \end{figure}

        \noindent
        In the first experiment, to verify the actual performance of the proposed DL-TPU, the trained DNN models for phase unwrapping with different high-frequency fringes are utilized to make predictions on the testing dataset (200 image pairs) (refer to Supplementary Section 2 for details), and MF-TPU is also implemented for comparison. In order to quantitatively analyze the accuracy of phase unwrapping for DL-TPU and MF-TPU, the phases with different high frequency are independently unwrapped by the two algorithms, and the average error rates for phase unwrapping on the testing dataset are calculated and plotted against ${f_h}$ in Fig. \ref{Fig2}${(a)}$. It should be noted that these results are calculated only by comparing the differences between the obtained phases and the label's phases for each valid point from the testing dataset (refer to Supplementary Section 2 for identifying the valid points). The label's phases can be correctly acquired as the `ground-truth' phase by exploiting multiple sets of phases with different frequencies to hierarchically unwrap the wrapped phase step by step. It can be seen from Fig. \ref{Fig2}${(a)}$ that with the increase of ${f_h}$ the reconstructed phases of MF-TPU are completely obviated, with a substantial increase of phase unwrapping error rate from 0 to ${12.71 \%}$. However, our approach always provides acceptable results, with more than ${95 \%}$ of all valid pixels being properly unwrapped. These experimental results confirm that compared with MF-TPU our method can achieve much better unwrapping results and decrease the phase unwrapping errors by almost an order of magnitude.

        In order to reflect the specific performance of DL-TPU and MF-TPU more intuitively, the 3D reconstruction results after phase unwrapping for a representative sample on the testing dataset are illustrated and compared in Fig. \ref{Fig2}${(b)}$, and the phase unwrapping error rates can be obviously seen in the background. It can be found from Fig. \ref{Fig2}${(b)}$ that our approach provides the smallest phase unwrapping errors and the significant improvement of phase measurement quality with the period number ${f_h}$ as expected. It can be further observed that the fringe order errors are mostly concentrated on the dark regions and object edges where the fringe quality is low. Different from MF-TPU, phase unwrapping errors caused by the low signal-to-noise ratio (SNR) region of phases is significantly reduced by using DL-TPU. For these low SNR region, the remaining phase errors have the characteristics of accumulation and can be easily further corrected by some compensation algorithm for fringe order errors \cite{zheng2017phase,zuo2018micro,yin2019high} (refer to Supplementary Section 4 for details of these compensation algorithms). Consequently, the trained models can substantially decrease error points to provide better phase unwrapping results (even ${f_h = 64}$) and lower error rates, which demonstrates the capability and reliability of DL-TPU for phase unwrapping.

    \subsection{Performance analysis under different types of phase errors}
        \subsubsection{Intensity noise}
            \noindent
            In the following a series of experiments, we will further verify the superiority of DL-TPU in the presence of different types of phase errors. In high-speed 3D measurement, the quality of the fringe patterns is poorer than that of the static measurement because it is projected and captured with limited exposure time. To emulate the practical measurement conditions, we measure a standard ceramic plate using DL-TPU (${f_h = 32}$) but artificially adjust the camera's exposure time to 39 ${ms}$, 20 ${ms}$, 15 ${ms}$, and 10 ${ms}$. To better analyze and compare the reliability of the accuracy results for phase unwrapping, the absolute phase map obtained using the 12-step phase-shifting algorithm and combining with a highly redundant multi-frequency temporal phase unwrapping strategy (with different frequencies including 1, 8, 16, and 32) can serve as the reference phase. Next, the error rate of phase unwrapping and the variance of the phase error ${\sigma_{\Delta \phi _{h}}}$ for different approaches are easily calculated by making a comparison between the unwrapped absolute phase and the reference phase for each valid point.

\begin{figure}[htbp]
            \centerline{\includegraphics[width=1\columnwidth]{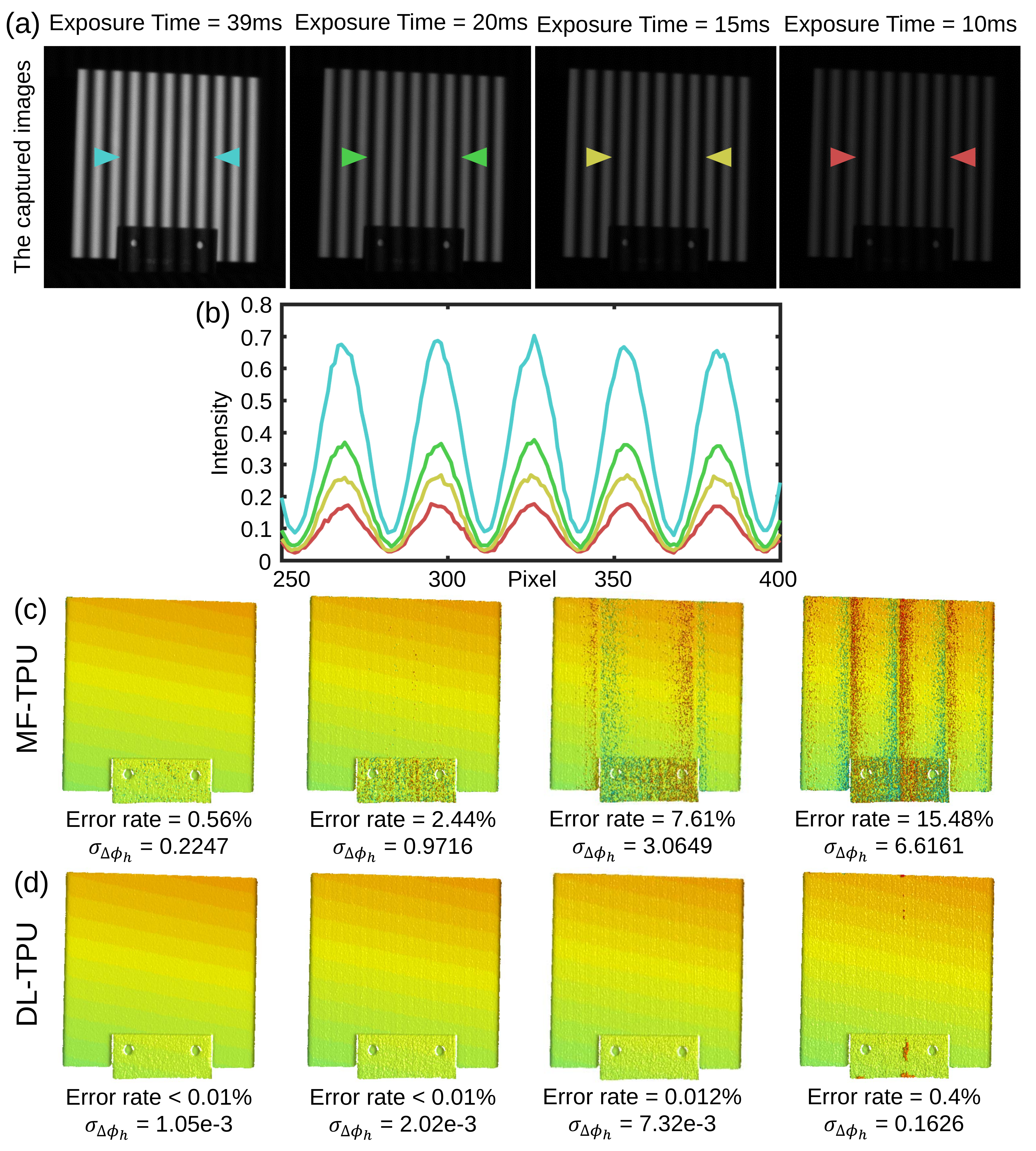}}
            \caption{
            (a) The captured images (${f_h = 32}$) of a standard ceramic plate under different exposure times. (b) Comparison of intensity in line 230 of the captured images. (c)-(d) Comparison of the 3D reconstruction results after phase unwrapping under different exposure times using MF-TPU and DL-TPU.
            }
            \label{Fig3}
            \end{figure}

            Obviously, as the exposure time decreases, the quality of the phase measurement drops significantly presented in Figs. \ref{Fig3}${(a)-(b)}$. Since the exposure time is a key factor affecting the speed and quality of phase measurement, the shorter the exposure time the algorithm can withstand, the faster the measurement can be achieved with six projection patterns in FFP. Therefore, a more robust phase unwrapping method is needed to eliminate the phase ambiguity introduced by reduced exposure times and make phase unwrapping correct. In Figs. \ref{Fig3}${(c)-(d)}$, it can be found that DL-TPU can always provide higher success rate of phase unwrapping and lower phase error ${\sigma_{\Delta \phi _{h}}}$ compared with MF-TPU, making it more appropriate for the high-speed 3D shape measurement applications.

        \subsubsection{Low fringe modulation}
\begin{figure}[htbp]
            \centerline{\includegraphics[width=1\columnwidth]{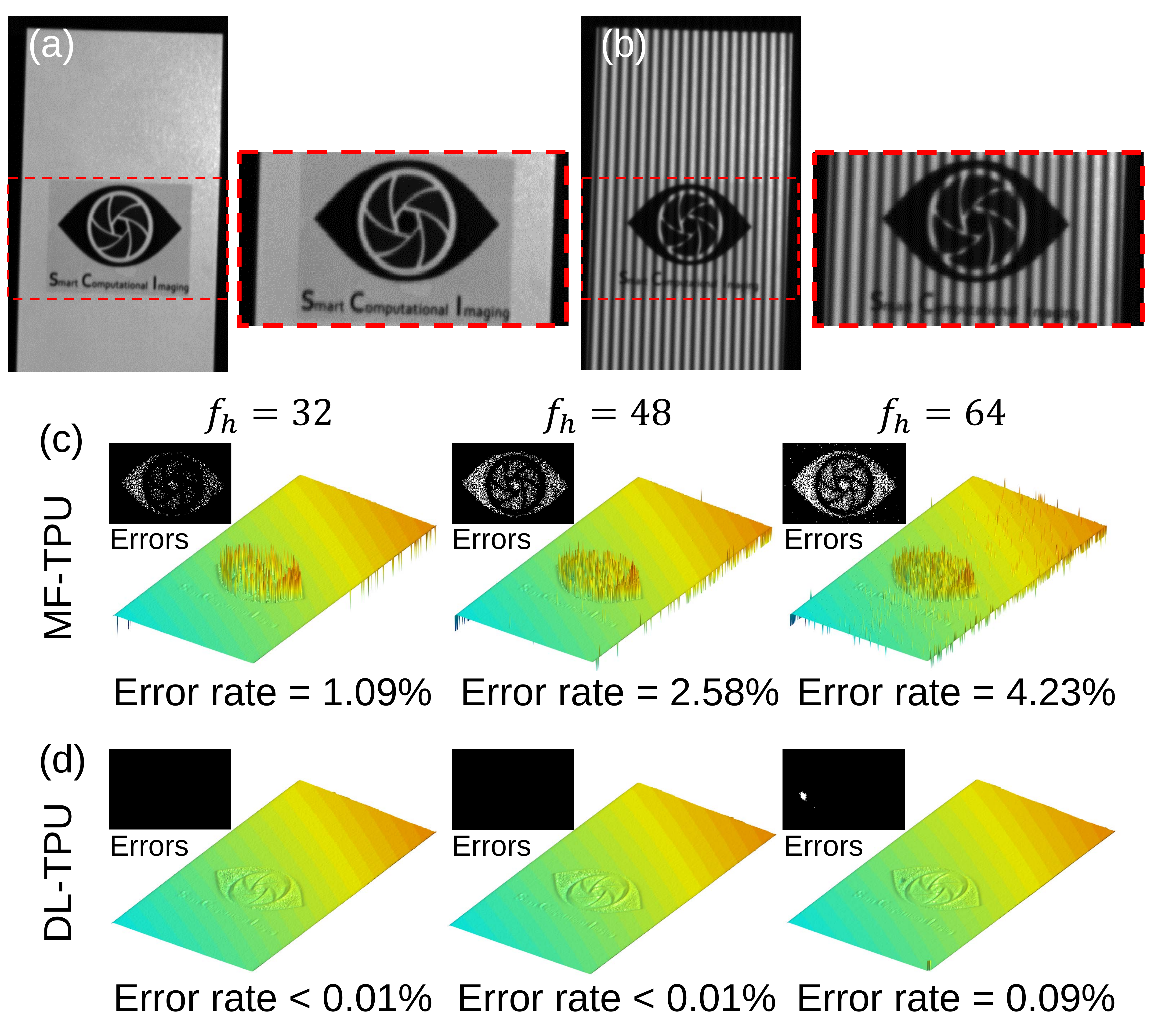}}
            \caption{
            (a) The tested object with the low-modulation logo. (b) The captured fringe image (${f_h = 64}$). (c)-(d)
            Comparison of the 3D reconstruction results after phase unwrapping for the low-quality region using MF-TPU and DL-TPU.
            }
            \label{Fig4}
            \end{figure}

            \noindent
            Another attractive attribute of DL-TPU is its good tolerance to noise that can significantly suppress phase unwrapping errors in low-fringe-modulation areas, which frequently appear in practical measurement for the surfaces of complex objects, like the tested object shown in Figs. \ref{Fig4}${(a)-(b)}$. For the low-modulation logo region, conventional MF-TPU results provide spinous results teemed with significant delta-spike artifacts, as shown in Fig. \ref{Fig4}${(c)}$. In contrast, the DNN approach successfully overcomes the low-SNR problem and produces smooth measurement results with negligible errors, as shown in Fig. \ref{Fig4}${(d)}$. This experimental result confirms once again that DL-TPU can provide superior capability and stability of phase unwrapping for suppressing unwrapping errors caused by low fringe modulation.

        \subsubsection{Intensity nonlinearity}
\begin{figure*}[htbp]
                \centerline{\includegraphics[width=2\columnwidth]{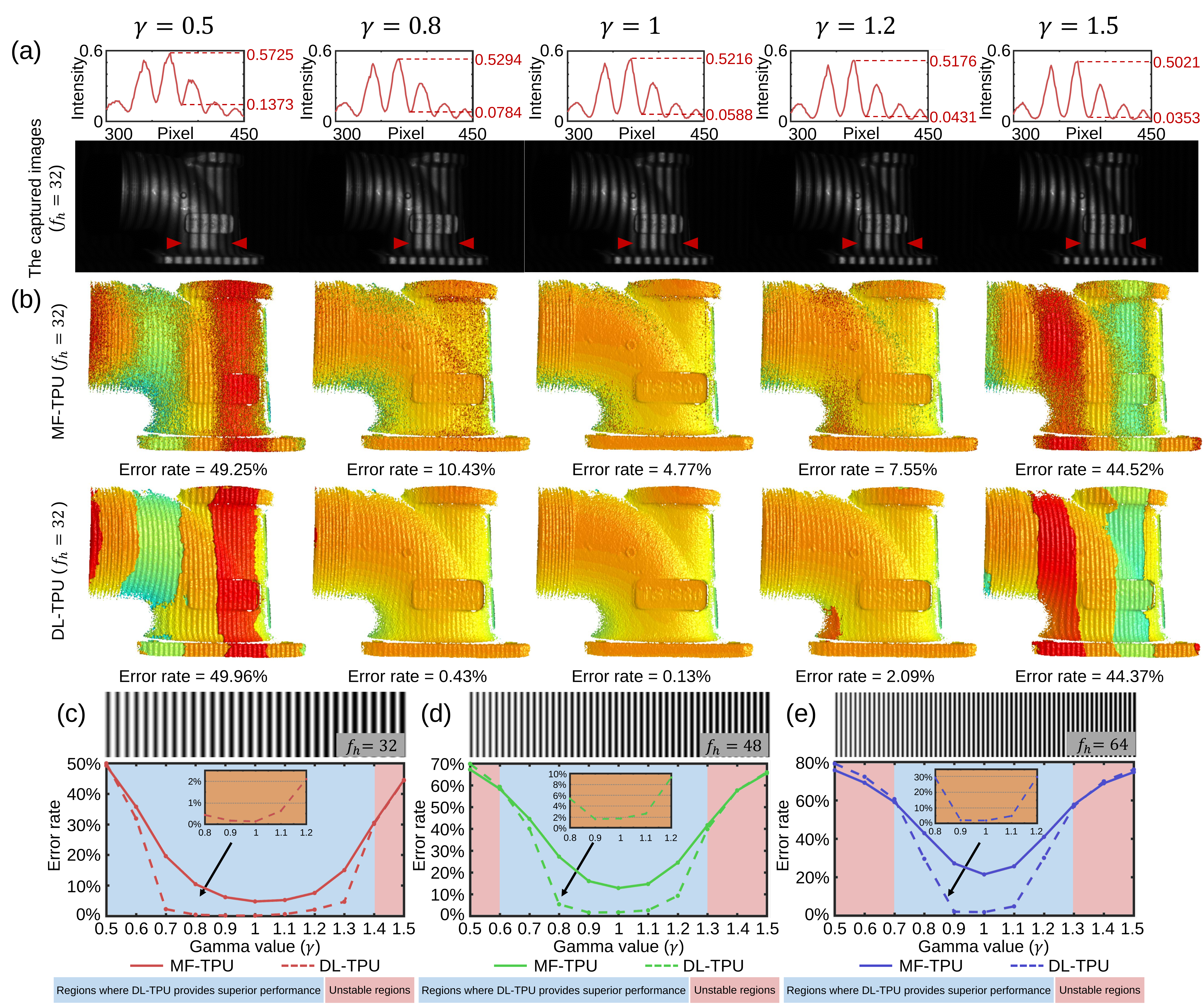}}
                \caption{
                (a) The captured fringe images (${f_h = 32}$) and the comparison of intensity in line 363 of the corresponding images under different degrees of intensity gamma distortion. (b) The 3D reconstruction results after phase unwrapping under different degrees of intensity gamma distortion using MF-TPU and DL-TPU when ${f_h}$ is 32. (c)-(e) The error rates of phase unwrapping with different high frequencies (such as 32, 48 and 64) under different degrees of intensity gamma distortion using MF-TPU and DL-TPU.
                }
                \label{Fig5}
            \end{figure*}

            \noindent
            In this section, we test the proposed DL-TPU under different degrees of intensity gamma distortion. The gamma distortion, or so called intensity nonlinearity, is a common error source in FPP due to the nonlinear response of the commercial projector, introducing high-order harmonics to the projected fringe patterns. The intensity of the fringes with the gamma distortion can be expressed as
            \begin{equation}\label{12}
            \begin{aligned}
                {I^{p,\gamma}_n(x^p,y^p)} = \left\{0.5 + 0.5\cos(2\pi f x^p - 2\pi n /3)\right\}^{\gamma},
            \end{aligned}
            \end{equation}
            where ${\gamma}$ is the gamma value that represents the nonlinear response of the commercial projector. Then, we choose an industrial workpiece of metal as the measured object to validate the resistance of DL-TPU to the gamma distortion. A set of fringe patterns with different gamma values, ranging from 0.5 to 1.5, are generated based on Eq. (\ref{12}) and projected onto the surface of the tested object in Fig. \ref{Fig5}${(a)}$. It can be seen from the 3D results shown in Fig. \ref{Fig5}${(b)}$ that MF-TPU cannot provide acceptable phase unwrapping results even under low-level gamma distortions. On the contrary, DL-TPU is able to achieve a close to ideal phase unwrapping result even when ${\gamma}$ is 0.8. It should be also noticed that, when ${\gamma}$ is as low as 0.5 or as high as 1.5, both of the two approaches can produce meaningful results since the phase errors artificially introduced is much larger than the "safe line" without triggering phase unwrapping errors, so that the success/error rate of unwrapping is about fifty-fifty. In Figs. \ref{Fig5}${(c)-(e)}$, for phase unwrapping with different high frequencies (such as 32, 48 and 64) under different degrees of intensity gamma distortion, the statistics curves of phase unwrapping for MF-TPU are shown as the solid lines, and the results are significantly improved by using DL-TPU as shown by dashed lines. These results verify that the proposed method can significantly reduce phase unwrapping errors and produce high-quality absolute phases even there is a certain degree of gamma distortion in the FPP system.

    \subsection{Application to high-speed 3D surface imaging}
\begin{figure}[htbp]
            \centerline{\includegraphics[width=1\columnwidth]{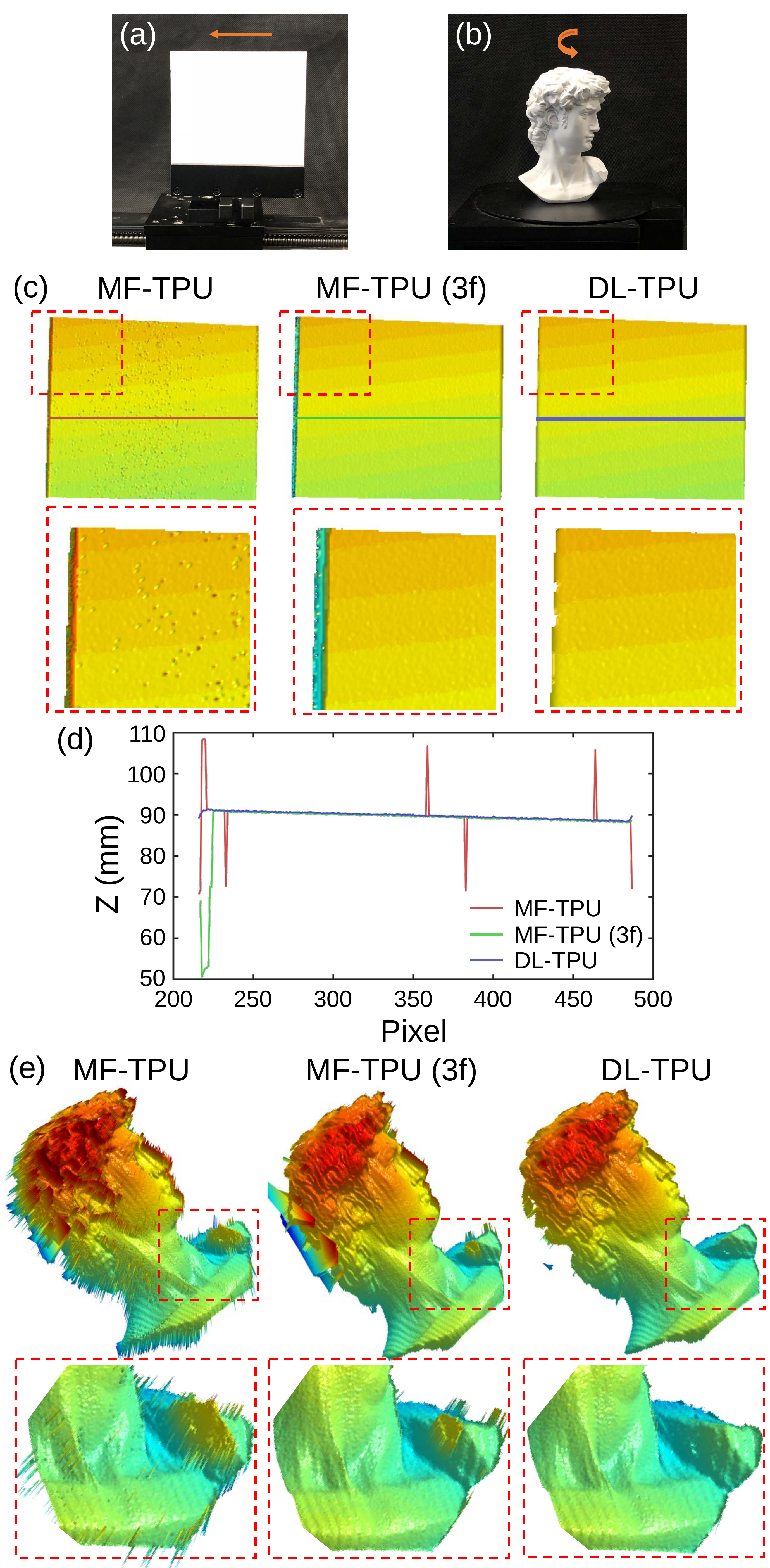}}
            \caption{
            (a)-(b) The objects with fast translation movement and rapid rotatory motion. (c) Comparison of the 3D results of phase unwrapping for the fast translation movement using MF-TPU, MF-TPU (3f), and DL-TPU. (d) The 3D result comparison in line 250 for the fast translation movement. (e) Comparison of the 3D results of phase unwrapping for the rapid rotatory motion using MF-TPU, MF-TPU (3f), and DL-TPU.
            }
            \label{Fig6}
        \end{figure}

        \noindent
        Finally, our system, which can project and capture the fringe images at the speed of 25 Hz, is applied to imaging some classical dynamic scenes for high-speed 3D shape measurement: objects with fast translation movement and rapid rotatory motion. In Fig. \ref{Fig6}${(a)}$, a standard ceramic plate, fixed on precision displacement platform, is performed to periodic translational movement at the speed of 1.25 ${cm/s}$. In traditional MF-TPU, it is more much difficult to unwrap the high-frequency phase using only one unit-frequency phase in Fig \ref{Fig6}${(c)}$ due to the unavoidable noises in actual measurement. Therefore, in order to ensure the stability of phase unwrapping for the high-frequency phase, three sets of phase-shifting fringe patterns, so-called MF-TPU (3f) in which the frequency of the second set of fringe patterns is 8, are used to achieve high accuracy but inefficient phase unwrapping. When measuring dynamic scenes, the motion will lead to phase distortion artifacts, especially when the object motion during the interframe time gap is non-negligible and more severe because of projecting additional patterns as presented in Fig. \ref{Fig6}${(c)}$. However, without the assistance of additional patterns, it illustrates the reliability and efficiency of DL-TPU from Fig. \ref{Fig6}${(c)}$ that the trained models can still achieve better phase unwrapping results. We try to take one cross-section on the 3D results of the ceramic plate to compare DL-TPU with MF-TPU and MF-TPU (3f). From the comparison results shown in Fig. \ref{Fig6}${(d)}$, it can be found that our approach provides the highest unwrapping reliability and best noise-robustness compared with other methods.

        And then, for measuring the rapid rotatory motion, the statue of David rotates in a counter-clockwise direction at the rotation rate of 3 ${rpm}$ as shown in Fig. \ref{Fig6}${(b)}$. Undoubtedly, in Fig. \ref{Fig6}${(e)}$, the experiment yielded a result similar to that of the fast translational motion. In the whole measuring procedures, the ceramic plate and the statue of David are correctly reconstructed with high quality, verifying the reliability of the proposed method to perform the absolute 3D shape measurement with high accuracy at high speed.

\section{Conclusions and Discussion}
    \noindent
    In this work, we have demonstrated the use of a deep neural network to significantly enhance the performance of TPU with high-frequency fringes acquired by a common FPP system. This high-performance TPU (so-called DL-TPU) can be achieved based on a deep neural network after appropriate training. Compared with MF-TPU, DL-TPU can effectively recover the absolute phase from two wrapped phases with different frequencies by exploiting both spatial and temporal phase information in an integrated way. It can substantially improve the reliability of phase unwrapping even when high-frequency fringe patterns are used. We have further experimentally demonstrated for the first time, to our knowledge, that the high-frequency phase obtained from 64-period 3-step phase-shifting fringe patterns can be directly and reliably unwrapped from one unit-frequency phase, facilitating high-accuracy high-speed 3D surface imaging with use of only 6 projected patterns without exploring any prior information and geometric constraint. After that, various experiments have been designed to access the phase unwrapping capability of the proposed approach under the conditions of intensity noise, low fringe modulation, and intensity nonlinearity. Experimental results have verified that TPU using deep learning provides significantly improved unwrapping reliability to realize the absolute 3D measurement for objects with complex surfaces. Besides, for the applications to high-speed FPP, it have also been observed that the deep learning-based approach is much less affected by motion artifacts in dynamic measurement and can successfully reconstruct the surface profile of the moving and rotating objects at high speed. These results highlight that machine learning is able to potentially overcome challenging issues in optical metrology, and provides new possibilities and flexibilities to design more powerful high-speed FPP systems. Although the TPU and FPP have been the main focus of this research, we envisage that the similar deep learning framework might also be applicable to other 3D surface imaging modalities, including, e.g., stereo vision \cite{lazaros2008review}, DIC \cite{pan2018digital}, spatial-temporal stereo \cite{zhang2003spacetime}, spatial-temporal correlation \cite{harendt20143d}, among others.

\section*{Funding Information}
    This work was supported by National Natural Science Foundation of China (61722506, 61705105, 11574152), National Key R$\&$D Program of China (2017YFF0106403), Final Assembly ``13th Five-Year Plan'' Advanced Research Project of China (30102070102), Equipment Advanced Research Fund of China (61404150202), The Key Research and Development Program of Jiangsu Province (BE2017162), Outstanding Youth Foundation of Jiangsu Province (BK20170034), National Defense Science and Technology Foundation of China (0106173), ``333 Engineering'' Research Project of Jiangsu Province (BRA2016407), Fundamental Research Funds for the Central Universities (30917011204), China Postdoctoral Science Foundation (2017M621747), Jiangsu Planned Projects for Postdoctoral Research Funds (1701038A).

	\bibliographystyle{IEEEtran}
	\bibliography{manuscript}

\end{document}